\documentclass[aip,reprint,floatfix]{revtex4-1}

\draft 

\usepackage[utf8]{inputenc}
\usepackage[T1]{fontenc}
\usepackage{textcomp}
\usepackage[english]{babel}

\newcommand{\dif}{\mathop{\text{d}}}

\usepackage{graphicx}

\usepackage{xcolor}
\usepackage[draft]{pdfcomment}
\setliststyle{AuthorComment}
\defineavatar{GH}{author={Gregor Hlawacek},color=yellow,opacity=0.5,voffset=5pt}
\pdfcommentsetup{avatar=GH}

\usepackage{cmap}

\hypersetup{
   pdftitle={Helium Ion Microscopy},
   pdfcreator={G. Hlawacek},
   pdfauthor={G. Hlawacek, V. Veligura, R. van Gastel, B. Poelsema},
   pdfsubject={$Revision: 899 $},
   pdfcreationdate={D:20130215132830},
   pdfkeywords={helium ion microscopy, gas field ion source, microscopy,
   nanofabrication},
   pdfstartview={FitH},
   bookmarks=false
}

\usepackage[all]{hypcap} 

\newcommand{\lowcite}[1]{[\onlinecite{#1}]}

\newcommand{\GH}[1]{{#1}}

\begin{document}


\title{Helium Ion Microscopy} 



\author{Gregor Hlawacek}
\email[Correspondence to: ]{g.hlawacek@utwente.nl}
\author{Vasilisa Veligura}
\author{Raoul van Gastel}
\author{Bene Poelsema}
\affiliation{Physics of Interfaces and Nanomaterials, MESA+ Research
Institute, University of Twente, PO Box 217, 7500AE Enschede, The
Netherlands}


\date{\today}

\begin{abstract}
   Helium Ion Microcopy (HIM) based on Gas Field Ion Sources (GFIS)
   represents a new ultra high resolution microscopy and nano--fabrication
   technique. It is an enabling technology that not only provides imagery of
   conducting as well as uncoated insulating nano--structures but also
   allows to create these features. The latter can be achieved using resists
   or material removal due to sputtering. The close to free--form sculpting of
   structures over several length scales has been made possible by the
   extension of the method to other gases such as Neon. A brief introduction
   of the underlying
   physics as well as a broad review of the applicability of the method is
   presented in this review.
\end{abstract}


\maketitle 


\section{Introduction}

High resolution imaging, in particular in biology and materials science, is
mostly done using Scanning Electron Microscopy (SEM). The ease of use and
the widespread availability has made this the number one method for
imaging in the aforementioned fields. Structuring and manipulation of
nano--sized features is traditionally the domain
of Focused Ion Beams. Here, mainly liquid metal ion sources (LMIS) using
Gallium are used. However, other techniques such as various types of
GFIS,\cite{Tondare2005} alloy LMIS,\cite{Bischoff2005} magneto optical
trap sources (MOTIS)\cite{Hanssen2008} and multicusp plasma
sources\cite{Ji2002a} are runners--up. Good reviews discussing these two
techniques can be found in Refs.~\lowcite{Goldstein2003} and~\lowcite{Utke2008} for SEM
and FIB, respectively. 

Helium Ion Microscopy presents a technique that unites many of the
advantages of SEM and FIB in a single tool. More importantly, it also
overcomes some of the deficiencies of SEM and FIB. In particular, the
possibility to image biological and in general insulating samples without
coating at highest resolution is important for many scientific questions
currently discussed in materials science and biology. Another important
charged particle beam technique---namely Transmission Electron
Microscopy (TEM)---depends on very thin samples free of defects from the
preparation. The unique nano--sized beam of the HIM makes it possible to not
only mill and pattern smallest features but also do this with minimal damage
to the crystal lattice.

In the following we will give an outline of the working principle and signal
generation in helium ion microscopy, followed by two sections highlighting
specific imaging applications and examples of materials modification.

\subsection{Working principle}

The initial idea of a scanning ion microscope has been put forward by Knoll
and Ruska already in 1932.\cite{Knoll1932} The working principle of a helium ion microscope can be divided into three
different stages. 
\begin{enumerate}
   \setlength{\itemsep}{1pt}
   \setlength{\parskip}{0pt}
   \setlength{\parsep}{0pt}
   \item Helium ionization and acceleration
   \item Beam formation and control
   \item Sample interaction
\end{enumerate}
In this introduction we will only briefly touch points 1 and 2. 
Instead, we will focus on the physics that is important to understand the
application of the technique to imaging and nanofabrication.

\begin{figure}[tbp]
   \centering
   \includegraphics[width=\columnwidth]{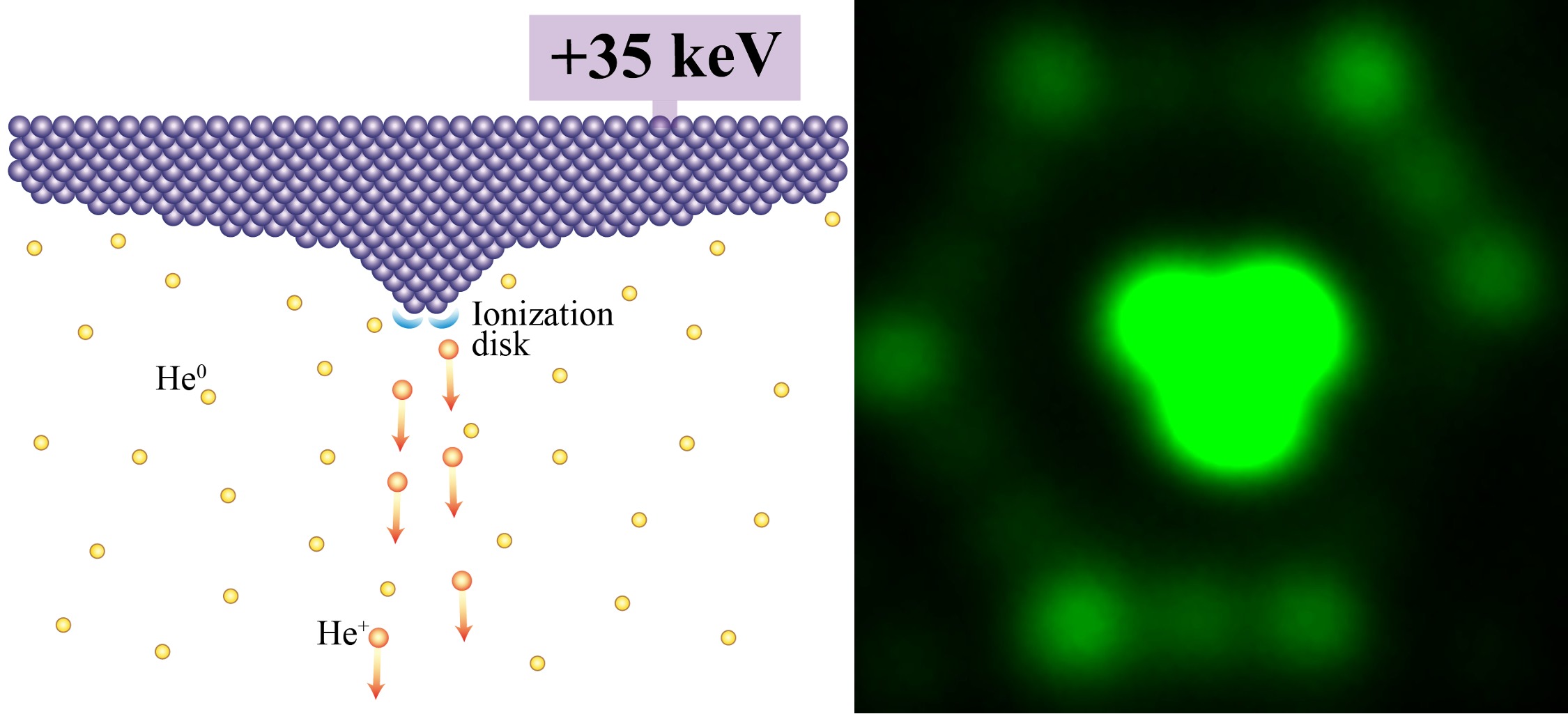}
   \caption{(Color online) (a) Sketch of a GFIS. Ionization happens dominantly at the
   most protruding corner and edge atoms. (b) Actual SFIM image of a GFIS.
The center trimer and the edges and corners of the next crystal plane are
visible.}
   \label{fig:HIM-source}
\end{figure}

The first is made possible by using a newly developed Gas Field Ion Source
(GFIS). GFIS have been investigated for a long
time\cite{Orloff1979,Tondare2005,Horiuchi1988,Sato1992} for their use in
microscopes.\cite{Escovitz1975,Orloff1975,Itakura1985,Horiuchi1988a} The
idea is based on the initial design of a field ion microscope by E.
M\"uller.\cite{Muller1951,Melmed1996}
However, only recently a stable ion source with a high brightness and small
virtual source size has been realized by Ward, Notte and Economou for use in a
microscope.\cite{Economou2006} It is based on an emitter whose apex has
been shaped into a three sided pyramid (see
Fig.~\ref{fig:HIM-source}). Work is done currently to understand and optimize
the supertip formation process in order to maximise the achievable
current.\cite{Kalbitzer2004,Pitters2013,Urban2012a,Urban2012,Pitters2012,Rahman2013}
Using Scanning Field Ion Microscopy (SFIM) the apex of the tip can be monitored and
shaped using high fields that can ultimately remove weakly bound atoms from
the apex. In this way the configuration of the tip apex can be controlled at
the atomic level. Although monomers are possible, trimers are more stable.
Figure~\ref{fig:HIM-source}(b) shows a SFIM
image of the tip. The trimer in the center and the edge of the first two
shelves below the trimer are visible. The combination of a pyramidally shaped
tip and the carefully shaped apex concentrates the electric field, so that
subsequent helium ionization predominantly happens at the top most atoms of
the pyramid. Using apertures in the beam path allows to select current
originating from ionization events on a single apex atom. Typical source
parameters are listed in table~\ref{tab:column-param}.

\begin{table}
   \centering
   \begin{tabular}{lcl}
      \hline
      Virtual source size & $\le$0.25\,nm & estimated\\
      Angular intensity & 0.5--1\,\textmu{}A\,sr$^{-1}$ & measured\\
      Brightness & $\approx1\times10^{9}$A\,cm$^{-2}$sr$^{-1}$ & calculated\\
      Energy spread & 1\,eV (0.25\,eV -- 0.5\,eV)\cite{Ernst1993} & measured\\
      \hline
   \end{tabular}
   \caption{Typical helium ion source parameters\cite{Hill2011}}
   \label{tab:column-param}
\end{table}

The second point is technologically demanding and requires a high degree of
knowledge on the design and implementation of the involved electrostatic
lenses, quadrupoles, octopoles, etc. For more details the reader
is refereed to numerous monographs available on charged particle
optics such as \lowcite{Orloff2003,Rose2009}. The critical source parameters important for the ion optical performance
of the column are given in table~\ref{tab:column-param}. The energy spread
of 1\,eV is an upper bound. Earlier measurements indicate that the values
could in fact be lower by a factor of two to four.\cite{Ernst1993} One of the
important consequences of the parameters listed in
table~\ref{tab:column-param} is the image side beam convergence angle
$\alpha_i$. Typical values for $\alpha_i$ are well below 1\,mrad. This small
beam divergence results in a large depth of field
\begin{equation}
   d_f=\frac{\delta}{\alpha_i}.
   \label{equ:DOF}
\end{equation}
Here, $\delta$ denotes the minimum feature that can be resolved in the
image. Assuming identical resolutions the HIM will have a superior depth of
field as compared to low--voltage SEM with typical $\alpha_i$ values of
8\,mrad.\cite{Hill2008}

Once the focused ion beam strikes the sample, the interaction of the
accelerated particles with the substrate atoms and electrons will give rise
to a large number of different signals. We will cover the available signals
in the following section. Before we do so, we will briefly discuss the
processes that occur during ion/sample interaction and that are important
for the achievable resolution in charged particle beam imaging. Besides the
small beam diameter, the shape and size of the actual interaction volume
between the impinging particle and the sample plays an important role to
reach ultimate resolution.
\begin{figure*}[tbp]
   \centering
   \includegraphics[width=\textwidth]{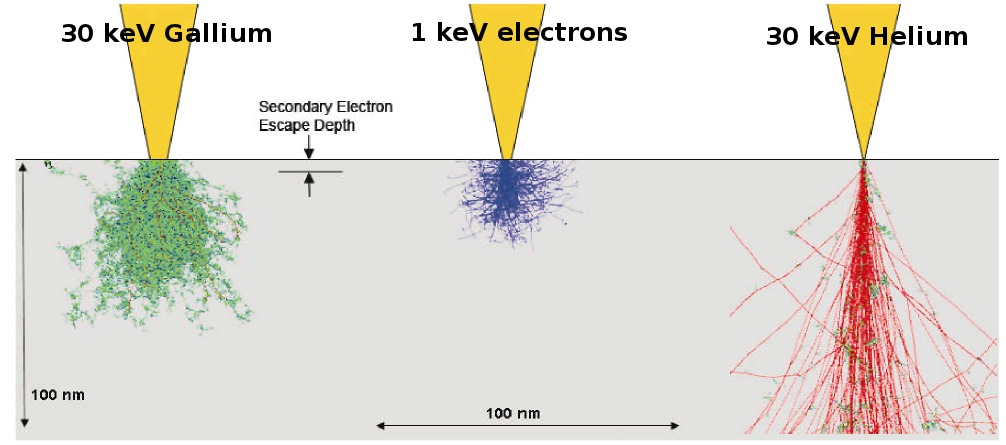}
   \caption{(Color online) Comparison of the interaction volume of different charged
      particles beams used for imaging. The substrate material is silicon in
      all cases. The typical escape depth of secondary electrons is
      indicated. From left to right, first the interaction volume of a
      30\,keV Gallium beam is shown. Such a beam is typically used in FIB
      applications. In the center a low energy electron beam of 1\,keV is
      presented. The energy is specifically chosen to maximize
      both surface sensitivity and resolution simultaneously. 
      Finally, a 30\,keV He beam---typically used in HIM to obtain maximum
      resolution---is presented. \GH{Ion trajectories have been
      obtained from SRIM.\cite{Ziegler2008} The electron trajectories have
      been calculated by CASINO.\cite{Demers2011}} Reprinted with permission
      from AIP Conf. Proc., Vol. 931, 489--496 (\citeyear{Notte2007}).
      Copyright (2007) AIP Publishing LLC.}
   \label{fig:charged-part-beams}
\end{figure*}
Figure~\ref{fig:charged-part-beams} compares Monte Carlo (MC) simulation
results for different charged particle beams. From the figure it is evident
that the interaction volume relevant for secondary electron (SE) generation
of the focused He beam is smaller than for the other two.\cite{Notte2007}
For the case of a Ga beam the large cross section of Ga with---in this case
Si---leads to substantial scattering in the near surface region relevant for
the SE signal generation. For a low energy electron beam---needed to
simultaneously optimize resolution and surface sensitivity in
SEM---electron--electron scattering in the sample widens the beam
dramatically in the first few nanometers deteriorating the achievable
resolution. Scattering also occurs in HIM. In the case of a 30\,keV He beam
scattering occurs with the nuclei of the sample atoms. However, due to the
low mass of helium, scattering is not very efficient in the first few
nanometers of sample material. This results in minimal beam divergence
inside the sample. Consequently the collected SE originate from a cylinder
with a minimal volume. The somewhat lower energy of the SE in
HIM\cite{Petrov2011} and the lower characteristic escape depth for SE in HIM
for light elements\cite{Ramachandra2009} enhances this difference between
SEM and HIM even further.

\subsection{Signals}

Next we will discuss available particles and correoponding signals. We will
in particular highlight their benefits and drawbacks when used for imaging
and what kind of physical quantities can be accessed using them.
The sequence in which they are discussed is determined by their abundance
in the tool. This also corresponds to the ease of use and how widespread the
technique is available in the current instrument base. An initial overview
of some of the different contrast mechanisms is given by
Bell.\cite{Bell2009,Scipioni2009a}

\subsubsection{Secondary electrons}

Secondary electron (SE) emission generated by ions can be split into two
parts. In the logical order we will first discuss SE generation followed by
the transport of electrons through the material. The latter is important as the generated
electrons still need to reach and subsequently leave the surface into the
vacuum so they can be detected.
As has been proposed by Bethe\cite{Bethe1941} the rate of secondary
electron generation $\delta_{SE}$ (electrons per incoming ion) is
proportional to the stopping power of the incident particle $\frac{\dif
E}{\dif s}$ in eV/\AA. 
\begin{equation} 
   \delta_{SE}=-\frac{1}{\epsilon}\frac{\dif E}{\dif s}
\end{equation}
Here, $\epsilon$ denotes a scaling constant. It is assumed that at least in
the relevant near surface region, atomic collisions do not play an important
role and $\frac{\dif E}{\dif s}$ depends only on the electronic stopping
power.

The generated SE1 are mostly excited by the incoming ions via kinetic
emission (KE).\cite{Ohya2008} Two types of secondary electrons of type 2
(SE2) exist in ion beam imaging. Type 2 electrons can be generated either by
recoiling target atoms or from other SEs of type 1. 
The second type of SE2 generation (SE generated by SE) does not play an
important role in HIM. This becomes clear when looking at the maximum energy
of the SE, which is taken to be equal to the energy loss of the impinging He for a
head--on collision\cite{Ohya2008,Rosler1985}
\begin{equation}
   \Delta{}E=2m_e\left( v+\left( v_F/2 \right) \right)^2.
   \label{equ:se_energy}
\end{equation}
Here, $m_e$ is the electron mass, $v$ the ion velocity, and $v_F$ denotes
the Fermi--velocity. The cross section for such a collision is highest if
the ion velocity---$v\approx1.3\times10^6$\,m/s for a primary energy of
35\,keV---matches the Fermi--velocity of the electrons in the material. For
gold and 35\,keV He this yields a maximum SE energy of 45\,eV. However,
this energy is approximately equal to---or even below---the effective energy
needed for SE generation by electrons in many materials.\cite{Lin2005} Thus
the size of the electron--electron collision cascade is restricted. 

However, kinetic excitation of electrons is also possible by recoil atoms,
provided they are fast enough so that their speed still matches the
Fermi--velocity of the target material. Electrons produced by recoiling target
atoms are usually called SE2. Ramachandra et al.\cite{Ramachandra2009}
calculated the ratio between SE2 and SE1 as function of energy and material.
The conclusion that can be drawn from their calculation is that for higher
primary energies a smaller SE2/SE1 ratio can be achieved for most materials
and consequently a higher resolution is possible.

The other process for electron emission is potential emission (PE) via Auger
neutralization. However, only for very low energies below 5\,keV PE becomes
dominant\cite{Ferron1981,Ramachandra2009} over KE. 

Once secondary electrons have been generated they still need to escape from
the solid. This process can be described as a diffusion process. The
characteristic length scale of this diffusion process---the effective
diffusion length of secondary electrons $\lambda_d$---is of the order of
1\,nm for nearly all materials.\cite{Ramachandra2009} This leads to the
fact that only the first few nanometers of the material add to the
emitted SEs. Measurements of the effective SE yield in HIM show
variations between 1 for carbon and values as high as 8 for
platinum.\cite{Notte2007}

\begin{figure}[tbp]
   \centering
   \includegraphics[width=\linewidth]{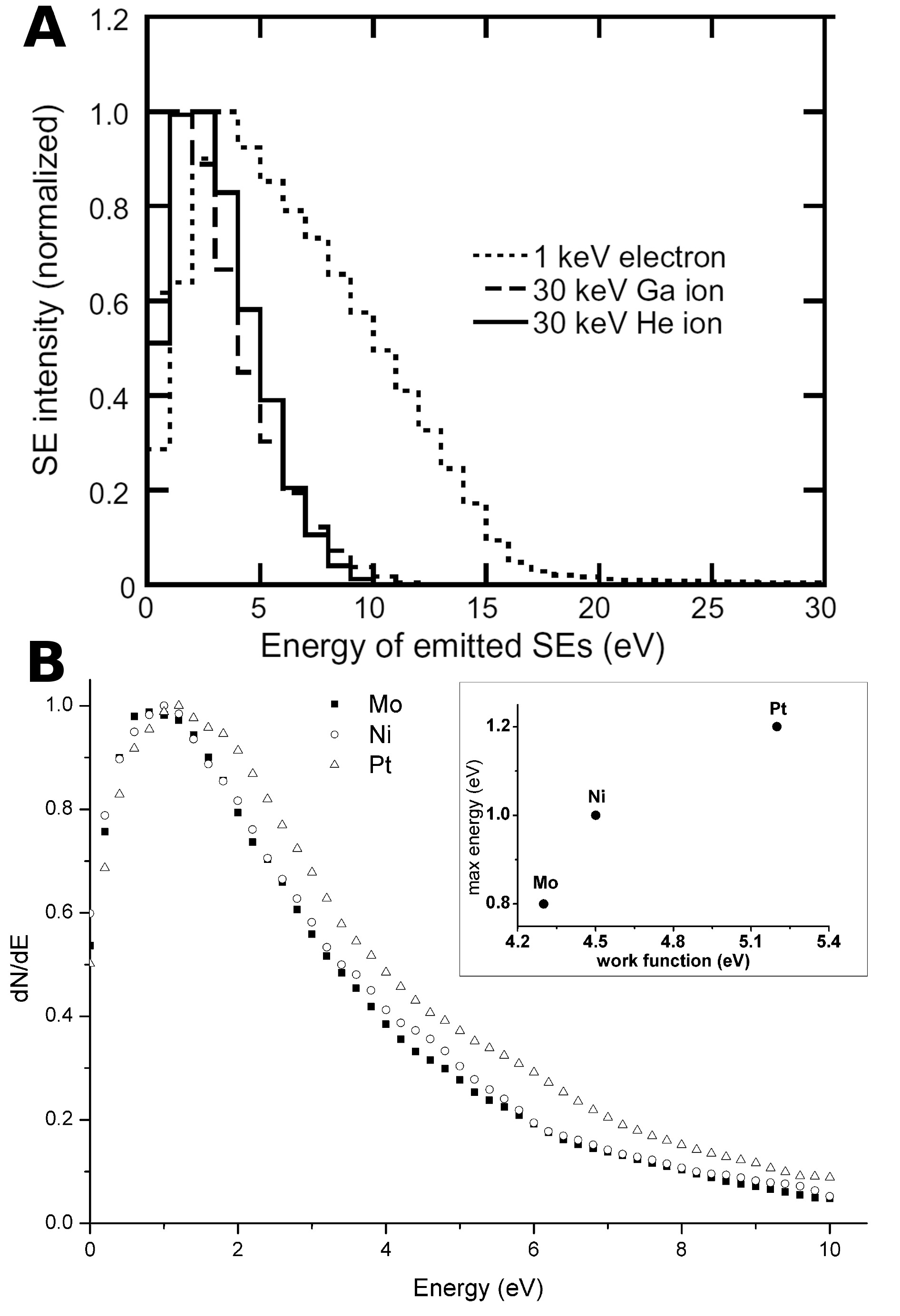}
   \caption{SE yield in HIM. (A) Comparison of calculated SE yields for
      electrons, Ga ions and He ions. Reproduced with permission
      from~\lowcite{Ohya2009}. (B) Experimentally obtained SE yield for three
      different metals. The dependence of the peak position on work function
      is shown as an inset. Reproduced with permission from Proc. SPIE, Vol.
      8036, 80360O-1--80360O-10 (\citeyear{Petrov2011}). Copyright (2011)
      SPIE.}
   \label{fig:SE-yield}
\end{figure}

The number and energy distribution of these ion induced secondary electrons
differs from what is found in a SEM. A sharper maximum at lower
energies is usually found\cite{Ohya2009,Petrov2010} in HIM. In
Fig.~\ref{fig:SE-yield}(a) a comparison of calculated SE yields in SEM,
Ga--FIB and HIM is presented. The calculations done
by Ohya et al. still overestimate peak position as well as peak
width.\cite{Petrov2011} These differences between actual measurement
results of SE yields in a HIM and calculations are attributed to SE
generation mechanisms not considered in the calculations. Indications exist
that bulk plasmon excitations can play an important
role\cite{Petrov2011,Riccardi2000} for SE generation in HIM. An actual
measurement of SE yield as a function of SE energy and the work function dependence of the
distribution maximum is presented in Fig.~\ref{fig:SE-yield}(b).
A consequence of the particular energy distribution of SE in HIM and the
small SE generation volume is an enhanced surface sensitivity. This has
been shown in a recent comparative study of HIM and SEM imaging performance
on carbon coated gold nano--rods.\cite{Hlawacek2013b} Only at very low
acceleration voltages is SEM able to visualize the thin carbon layer
covering the gold particles. However, low voltage SEM suffers from carbon
deposition in the imaged area making detailed studies challenging. An
example of the high sensitivity of HIM with respect to the material work function
is the visualization of different half unit cell surface termination in
Ti$_3$SiC$_2$.\cite{Buchholt2011} Depending on the position of the cleavage
plane the surface is either terminated by Si (half unit cell) or Ti (full
unit cell). The difference in the chemistry of the top surface layer of
atoms results in different SE yields for the two terminations. As a result
they can be discriminated in HIM.

A software package called \emph{IONiSE} developed by P. Rack and coworkers is available
that allows the calculation of the expected SE yield for a large number of
materials.\cite{Ramachandra2009} Although good agreement has been achieved
between simulation and experiment, only a limited number of analytical
applications based on SE yield are known.\cite{Petrov2011}

Sample topography in HIM is made visible in a similar way as in SEM. The
dependence of the SE yield on the specimen tilt with respect to the
incoming beam can be described the following secant law
\begin{equation}
   \delta_{SE}(\theta)=\delta_{SE}(0)\sec \theta.
   \label{equ:se-angle}
\end{equation}
However, experimental studies showed that the actually measured SE yield at
the detector deviates from the expected secant behaviour.\cite{Bell2009}
The deviations lead to a more linear response curve, which in turn should
reduce the strong edge effect known from SEM. Nevertheless, a very strong edge
effect has been observed in thin layers.\cite{Behan2012,Fox2013} 

In summary, the achievable high resolution and surface sensitivity in HIM is
based on the fact that the SEs originate from a cylinder at the beam
penetration point with a diameter of approximately 1\,nm---given by the beam
diameter---and a length of less than 3\,nm---determined by $\lambda_d$. This
small probe volume helps to
achieve very high image resolution in HIM. It should be noted that obtaining
such high resolution images has become substantially easier since the
introduction of UHV HIM.\cite{vanGastel2011} At small fields of views
usually high fluences are reached as a consequence of the large pixel
density. In an UHV HIM implantation and sputtering can still negatively
affect the sample during imaging. Carbon deposition on the other hand can be
excluded.\cite{Veligura2012} The removal of hydrocarbons from the sample
chamber vacuum prevents the formation of carbon deposits in the imaged area.
As a consequence some exceptional imaging results could be achieved (see
Fig.~\ref{fig:bio}(b)).

\subsubsection{Backscattered helium}

\GH{What backscattered electrons are to the SEM, backscattered helium (BSHe)
atoms and ions are to HIM.} This rather bold statement is true in several ways, as
will be highlighted in the next paragraphs.

Two different ways to utilize BSHe are available in current HIM. First, and most
commonly used, is a microchannel plate (MCP) detector to acquire qualitative element
distribution maps. Second, a silicon drift detector can be used to obtain
spectroscopic information from microscopic areas. The latter allows 
quantitative element identification based on the same principles as
Rutherford backscattering spectroscopy (RBS).

\begin{figure}[tbp]
   \centering
   \includegraphics[width=\columnwidth]{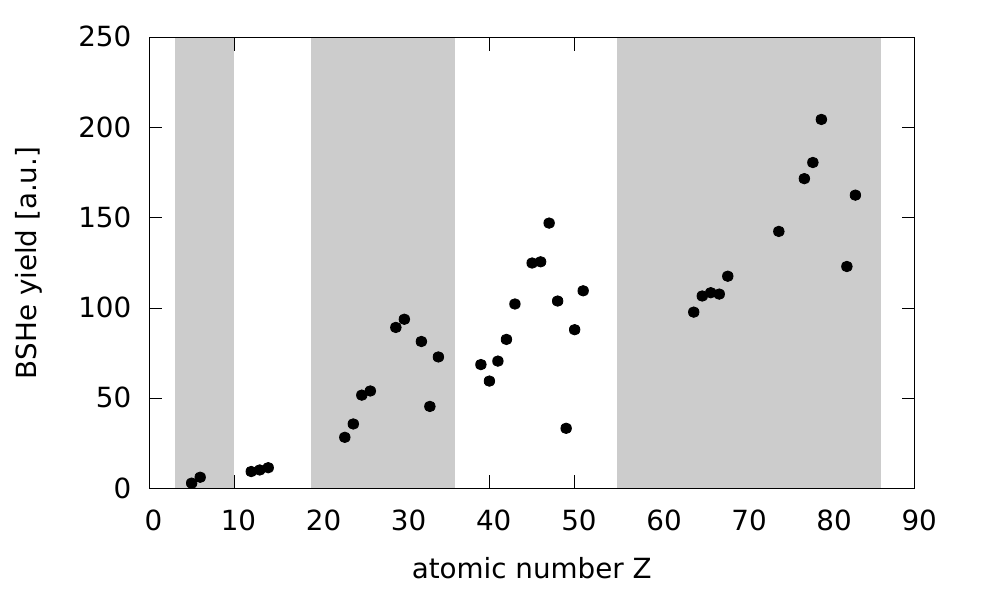}
   \caption{Experimentally obtained backscattered Helium yield plotted against atomic number
   $Z$. The row structure of the periodic table can clearly be seen on top
of a general increase in BSHe yield for heavy elements. Data courtesy of
Carl Zeiss AG.}
   \label{fig:BSHe-yield}
\end{figure}

While SE images usually are rich in morphological contrast, BSHe images
obtained with the MCP are poor in topography and rich in elemental contrast.
In contrast to the SE images, the information in these images originates
from the bulk of the sample. To measure them the MCP is inserted below the
pole piece in the primary beam path. While a center hole allows the primary
beam to reach the sample, this geometry maximizes the solid angle, and thus
the collection efficiency of the detector. The obtained contrast can be
understood by examining the Rutherford scattering cross section
\begin{equation}
   \dif\sigma=\left( \frac{q^2Z_1Z_2}{4E_0}
   \right)^2\frac{\dif\Omega}{\sin^4\frac{\theta}{2}},
   \label{equ:cross-section}
\end{equation}
where $q$ is the elementary charge, and $Z_1$ and $Z_2$ denote the atomic
number of impinging and target particle, respectively. Assuming that the
target atom is at rest, $E_0$ is the energy of the impinging particle,
$\dif\Omega$ is an arbitrary element of solid angle, and $\theta$ is the
scattering angle. For fixed $Z_1$---helium for this review---and a given
energy, a dependence on $Z_2^2$ leads to a strong contrast between different
elements. A more detailed analysis of the scattering cross section shows
that a dependence that is related to the structure of the periodic table of
elements also exists. This is a result of the change in screening along the
rows of the periodic table. In Fig.~\ref{fig:BSHe-yield}, experimentally
obtained BSHe yields for various elements are presented.
The dependence of the yield on the structure of the periodic table is
clearly present.

At this point it is important to realize that a high sensitivity is needed
from the detector that is used. Under standard imaging conditions in HIM,
typically 500 or less ions are used per pixel. Assuming a high backscatter
yield of 20\% and taking into account the detector solid angle, no more than
50 ions will reach the detector. 
A large fraction of these backscattered He particles are also
neutral\cite{Behan2010} due a charge transfer process which occurs once they
enter the sample. 
However, this would represent an ideal case for a heavy element bulk target.
In practice these numbers can be substantially smaller and not a single ion
should be lost in the detector.\cite{Hlawacek2013b}

One should realize that contrary to the electrons discussed in the previous
section, BSHe represent a bulk signal. Depending on the atomic number
sampling depth, between a few tens of nanometers and a few hundred nanometers
can be achieved. As can be seen from equ.~(\ref{equ:cross-section}), the
backscatter efficiency can be increased by lowering the acceleration
voltage. As a side effect, this will also reduce the range of the helium and
thus the sampling depth.

In addition to the qualitative element distribution, also quantitative
information on the elemental composition can be obtained. For a binary
collision---when momentum and energy are conserved---the kinetic energy of
the backscattered helium
\begin{equation}
   E_1=E_0\left( \frac{M_1}{M_1+M_2} \right)^2\left(
   \cos\theta\pm\sqrt{\left( \frac{M_1}{M_2}\right)^2-\sin^2\theta }
   \right)^2
   \label{equ:energyloss}
\end{equation}
depends on the on the ratio of the masses of the impinging ($M_1$) and
target ($M_2$) particle. Measuring the energy loss at a fixed angle allows
the mass of the collision partner, as well as its position relative to the
sample surface, to be determined. At high keV or low MeV energies this is
known as Rutherford Backscatter Spectroscopy (RBS). This method is known to
deliver high sensitivity and excellent depth-resolution. The nature of the
involved square function makes the method very sensitive to differences
between light elements. A silicon drift detector with a resolution of
approximately 4\,kV can be used for this purpose. However, due to the
relatively low
primary energy of only 35\,keV, the measured peaks are not as well defined
as in classical RBS. Nevertheless, the nature of HIM allows such spectra
from areas of only several \textmu{}m$^2$ to be obtained.\cite{Behan2012}
The technique has been successfully applied to measure ZrO$_2$
layer thicknesses on Si substrates with monolayer
sensitivity\cite{Sijbrandij2010} and for thickness measurements of
cobalt/nickel nano--rings.\cite{Behan2012}

The achievable resolution is not only limited by the detector. As a result
of the already low primary energy, the backscattered particles have a high
probability to undergo multiple scattering events. This occurs for the
impinging, as well the backscattered helium particle. Consequently, a
statistical broadening of the exit energies of the BSHe exists. It has been
shown by simulations that even for an ideal combination, such as a 75\,nm
thick heavy hafnium oxide film on silicon, an intrinsic uncertainty of 14\%
exists for the depth resolution of Hf. Given the current detector resolution
this value is further increased to 60\%.\cite{Gastel2013} Compositional
characterization will have even larger errors due to a severe peak
overlap.\cite{Behan2012}
\begin{figure}[tbp]
   \centering
   \includegraphics[width=\columnwidth]{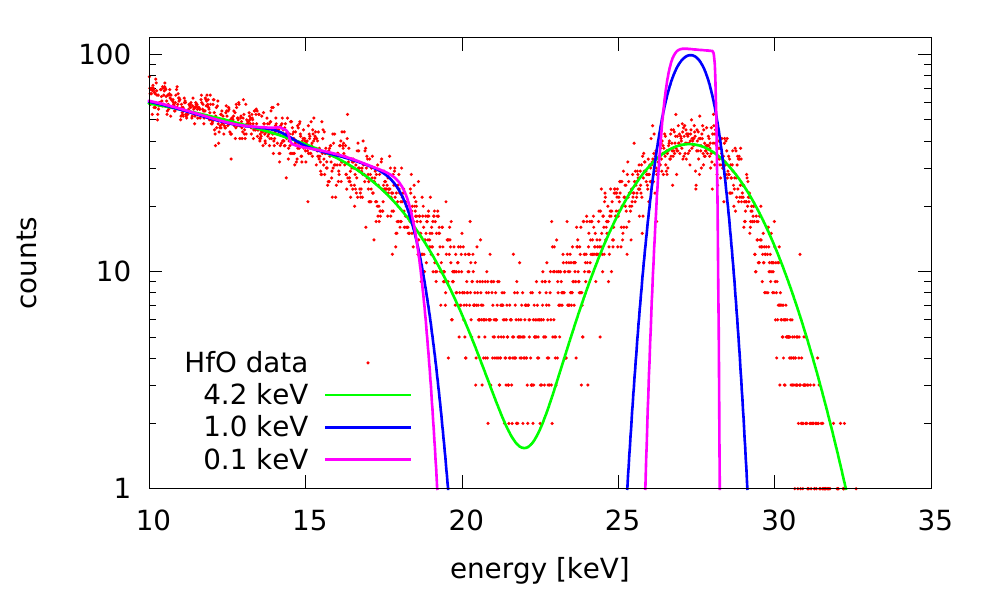}
   \caption{(Color online) Experimentally obtained RBS data from a 4.2\,nm hafnium oxide
   thin film on Si. The best fit to the data with the current used silicon
drift detector and estimated profile shapes with detectors of 1\,keV, and
100\,eV resolution are presented. }
   \label{fig:HfO}
\end{figure}
Figure~\ref{fig:HfO} presents an experimentally obtained spectrum from
hafnium oxide. The SIMNRA\cite{Mayer1999,Sijbrandij2010} calculated fits to
obtain elemental composition and layer thickness have been added for
different real and ideal detector resolutions. An increase of detector
resolution to 1\,kV would substantially improve the quality of the obtained
spectra and consequently of the fit accuracy. These curves should be taken
to be an indication of what could be possible and not an exact
representation of the achievable resolution. One also needs to keep in mind
that the majority of the backscattered particles are
neutral,\cite{Behan2010,Arafah1989} complicating any post scattering
treatment.

The yield of both signals---secondary electrons as well as back scattered
helium---also depends on the crystalline orientation of the sample with
respect to the beam. This allows for two additional contrast mechanisms
in HIM. The well known channeling
contrast\cite{Bell2009,Petrov2010} can be exploited to obtain the crystal
orientation of samples with a large lateral resolution.\cite{Veligura2012}
This technique makes use of calculations of the blocked area fraction, which
yields results similar to stereographic projections of channeling minima or
Laue back reflection patterns.

One of the surprising results is that despite the very small wavelength of
He ions, Scanning Transmission Ion Microscopy (STIM) is possible. Both dark field
and bright field images could be recorded in which thickness fringes and
line defects could be identified.\cite{Notte2010}

\subsubsection{Photons}

Generally speaking, ionoluminescence (IL) is a phenomenon of light emission due to
the optical transitions of an electronic system which has been excited by
ion irradiation. Three main stages can be distinguished in the luminescence
process:\cite{Lum_Marf,Gotze2009}
\begin{enumerate}  \itemsep0pt \parskip0pt
   \item energy absorption and excitation of the electronic system;
   \item system relaxation and energy transfer to the emission centers;
   \item transition of the system into the ground state by photon emission.
\end{enumerate}

During sample bombardment with He$^{+}$ ions, light can be
obtained from the excited backscattered neutral
He,\cite{He_emission1,HE_emission2} excited sputtered atoms and molecular
complexes,\cite{Sput_emission1,Sput_emission2,Sput_emission3} or from the
sample material itself.\cite{Sput_emission2,Sput_emissionBook}
For the case where emission originates from excited species which have left the
sample, the spectrum consists of discrete
Doppler--broadened lines. These lines corresponds to the optical transitions within
atomic (or molecular) orbitals. This light is usually observed at a distance
up to a few millimeters from the target surface.\cite{Sput_emission2} In
the case of
organic samples, ionoluminescence reveals the
electronic structure of the organic compounds.\cite{Bio_Pallon}

For the investigation of material properties we are mainly interested in the
luminescence from the sample itself. The physics of the emission processes is
usually described by considering the sample's electronic band
structure, or by using configuration coordinate diagrams.\cite{Lum_Marf}
According to its origin, there are two general types of luminescence:
\textit{extrinsic} and \textit{intrinsic}.\cite{German_book} In the case
of extrinsic luminescence, the light generation is related to the presence
of impurity atoms or ions (so--called \textit{activators}) in the sample
material. Depending on their electronic structure, activators can be
divided in the following groups:
\begin{enumerate}  \itemsep0pt \parskip0pt
   \item transition metal ions with d$^{n}$ electronic configuration (e.g. Ti$^{3+}$, Cr$^{3+}$, Mn$^{2+}$);
   \item ions with s$^{2}$\,-\,configuration (e.g. Tl$^{+}$, Pb$^{2+}$, Sb$^{3+}$);
   \item rare--earth elements (REE$^{2+/3+}$);
   \item actinides (e.g. UO$_{2}^{2+}$, Cm$^{3+}$).
\end{enumerate}
Sometimes the presence of a certain impurity (\textit{sensitizer}) is needed
for the luminescence of an activator (for example Ce$^{3+}$ for activation
of Tb$^{3+}$ ions~\cite{Hoffman1971}). 
As opposed to sensitizers, \textit{quenchers} suppress light emission from
an activator. 
For example, Fe$^{2+}$ ions act as quenchers for emission from Mn$^{2+}$ in
apatite.\cite{Ion_quench} 
At high activator concentrations self--quenching may occur due to resonant
absorption processes.

Intrinsic luminescence on the other hand, is emission from the sample
material itself. Two cases can be distinguished. First, optical transitions
from delocalized states, or in other words the recombination of
free electrons from the conduction band with holes in the valence band. This
can 
include direct and indirect transitions. The radiative recombination of
free excitons also falls in this category. Second, optical transitions from localized states can also be
attributed to intrinsic luminescence. This includes the following
processes: recombination of excitons trapped at defect sites
(so--called self--trapped excitons\cite{STE}), emission from excited
defects---known as color centers (e.g. nitrogen--vacancy centers in
diamond\cite{CC_diamond}), and transitions of charge carriers from delocalized
into localized states. The shape and width of the emission peaks and
bands strongly depends on electron--phonon interactions and thermal
effects. A strong electron--phonon interaction leads to a Stokes shift and
peak broadening. As a result, it is desirable to perform ionoluminescence
measurements at cryogenic temperatures, which has not been done to date in HIM.

Since ionoluminescence is in many aspects similar to the cathodoluminescence
(CL) phenomenon often observed in scanning electron microscopy (SEM),
databases of CL studies can be employed for the interpretation of IL
spectra. Ionoluminescence studies are significantly complicated by the fact
that an ion beam not only induces light emission, but also directly
influences the optical properties of the target due to defect creation. Ion
irradiation can lead to target coloration (e.g. in alkali
halides\cite{IC_color}) and enhanced emission, but also quench the
luminescence (e.g. semiconductor materials\cite{IL_Boden}). However, the use
of HIM to observe IL phenomena allows these processes to be followed
in--situ with a high lateral resolution.

\section{Microscopy}

In this section we will try to give an overview of applications of helium
ion microscopy that
make use of the special imaging capabilities of the HIM. Although, the high
resolution is the most prominent fact that allows for very accurate
critical dimension measurements,\cite{Postek2011} many successful applications of HIM
make use of other distinct characteristics of HIM. This overview is by no
means complete but will attempt to highlight interesting and eventually surprising
imaging applications. 

\subsection{Insulating and biological samples}

\begin{figure}[tbp]
   \centering
   \includegraphics[width=.8\linewidth]{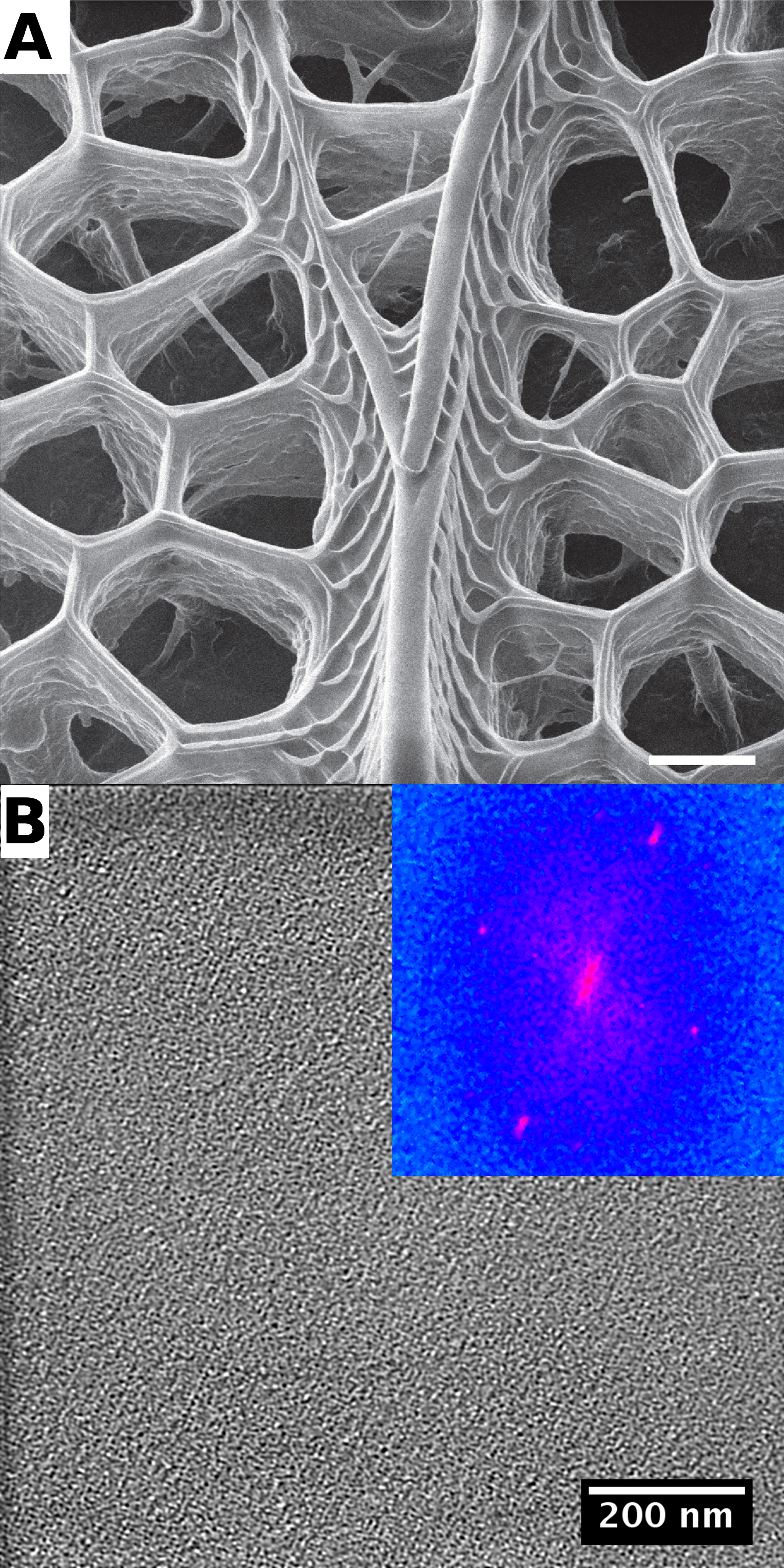}
   \caption{(Color online) HIM imaging of biological samples. (a) High resolution image of
      uncoated Papilio ulysses black ground scales. The scale bar is 400\,nm.
      \citetext{Reproduced with permission from Scanning 34, 107--20 (\citeyear{Boden2012a}). Copyright 2012 Wiley
      Periodical, Inc.} (b) High resolution HIM image of flat bovine liver
      catalyse crystal. The inset FFT
      highlights the resolved lattice spacings. Image obtain on a UHV-HIM courtesy
   of A. Lysse, Carl Zeiss Microscopy}
   \label{fig:bio}
\end{figure}

The use of electrons for charge neutralization enables HIM to obtain high
resolution images of insulating, and in particular uncoated, biological
samples. In Fig.~\ref{fig:bio}(a) a high resolution HIM image of a butterfly
wing is presented.\cite{Boden2012a} The black ground scales of Papilio
ulysses and other butterflies are imaged without any prior coating, which
allows the smallest features to be identified. The large depth of field in
HIM also makes it ideal for creating anaglyphs using different angles for
imaging. High precision measurements of otherwise not accessible feature
heights are possible this way.\cite{Boden2012a}
However, only very few groups have demonstrated the applicability of HIM for
imaging of cells.\cite{Scipioni2009a,Jiang2010,Joens2013} This is astonishing since
very high resolution should be possible. Figure~\ref{fig:bio}(b)
demonstrates the achievable resolution on such biological, soft and
insulating samples. The imaged protein crystal (flat bovine liver catalyse)
exhibits a simple rectangular unit cell with a lattice spacing of 8.8\,nm
$\times$ 6.7\,nm. These values are in excellent agreement with the expected
values from literature.\cite{Luftig1967}

\subsection{Subsurface imaging}

\begin{figure}[tbp]
   \centering
   \includegraphics[width=\linewidth]{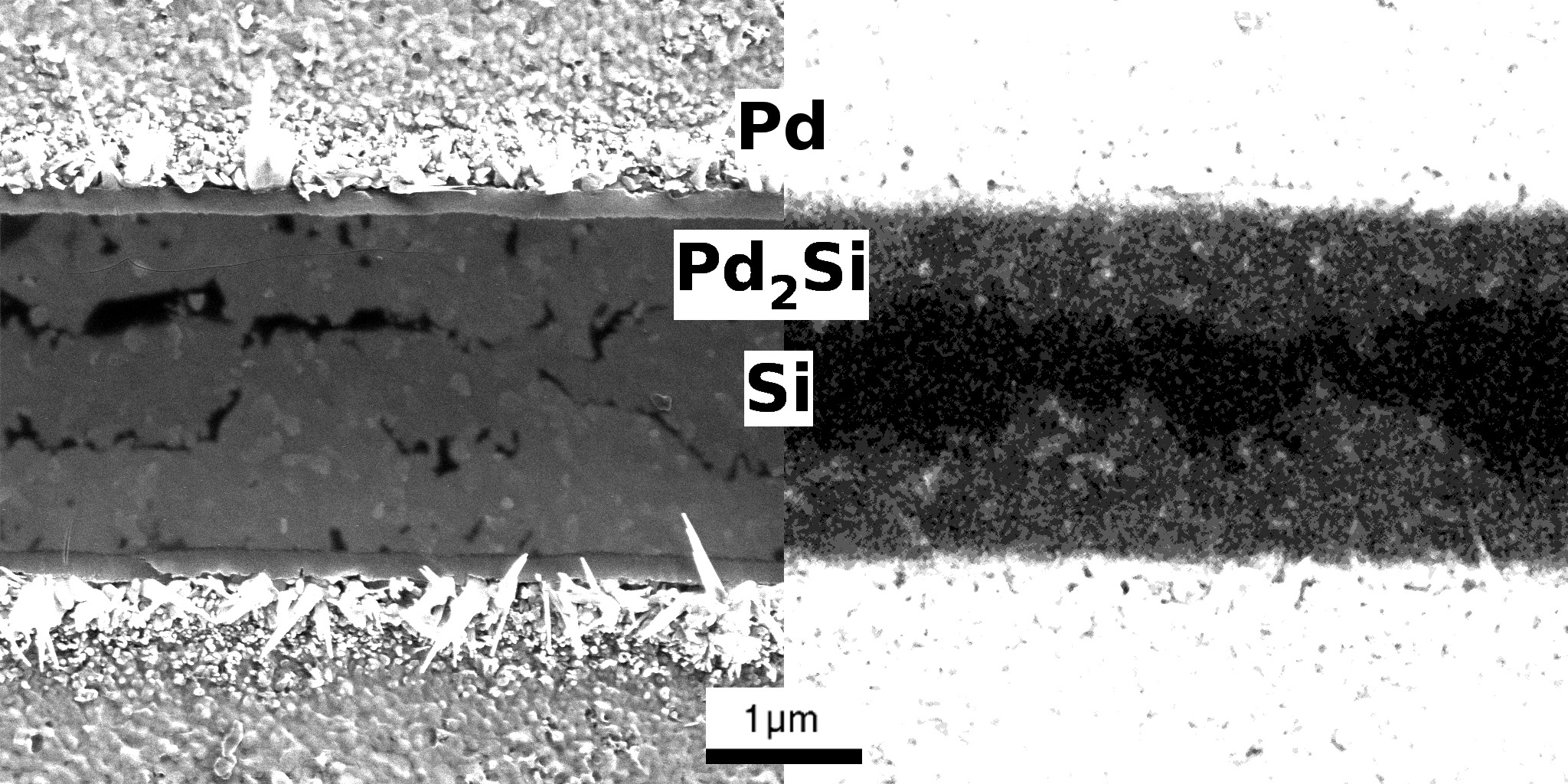}
   \caption{Subsurface imaging in HIM. (a) Surface sensitive SE image
      obtained from SiO$_2$ covered Pd$_2$Si interconnects. (b)
      Simultaneously recorded BSHe image showing the Pd deposits (bright,
      top and bottom) and the buried Pd$_2$Si layer. The shape of the two Pd
   diffusion fronts is clearly discernible in between the two Pd contacts.}
   \label{fig:buried}
\end{figure}

The use of BSHe for imaging allows amongst other things subsurface processes
such as the formation of buried contacts to be visualized. This overcomes an
existing limitation in how we currently try to follow subsurface diffusion
processes. Helium ion microscopy offers a unique, destruction free method to
reveal the in--plane shape of a diffusion front. In Fig.~\ref{fig:buried}
results of a HIM study on the subsurface formation of Pd interconnects are
presented.\cite{Gastel2012}
While a large degree of surface detail is present in the SE image presented
in Fig.~\ref{fig:buried}(a), no direct indication of the buried Pd$_2$Si
interconnect is visible. However, from the simultaneously recorded BSHe
image shown in Fig.~\ref{fig:buried}(b) the morphology of the Pd diffusion
front can be seen. The benefit of this method over cross section approaches
is evident. Besides the obvious ease with which the shape of curved
interfaces can be resolved, sample preparation is also substantially easier.
Knowledge of the non--straight nature of the diffusion front is important to
understand the reason for large variations in device performance. The shape
of the subsurface diffusion front is not accessible using cross section
techniques. Because the penetration depth of the ions can be varied by using
different primary energies, it is also possible to get an estimate of the
thickness of the cover layer.\cite{Gastel2012}

\subsection{Ionoluminescence}

There are several reviews\cite{Sput_emission2, Townsend2007, Townsend2012}
introducing various applications of IL for material characterization that 
also reveal possible complications of the technique. Here, we will focus
on IL imaging with light ions and results obtained with HIM. 

The possibility to do IL imaging using a HIM was demonstrated by Boden et
al.\cite{IL_Boden} The authors tested a wide range of samples such as
quantum dots (QD), semiconductors, rare--earth doped nanocrystals, and
Ce$^{3+}$ doped garnet. Both bulk samples and nanocrystals doped with REE
were found to exhibit IL. This is a result of the optical transitions of the
$f$ electrons in the dopant. However, the IL signal was found to decay with
prolonged beam exposure. This decay has been attributed to defect accumulation
which hinders electron transport to the actual emitting rare--earth atoms.

\begin{figure}[tbp]
   \includegraphics[width=.8\linewidth]{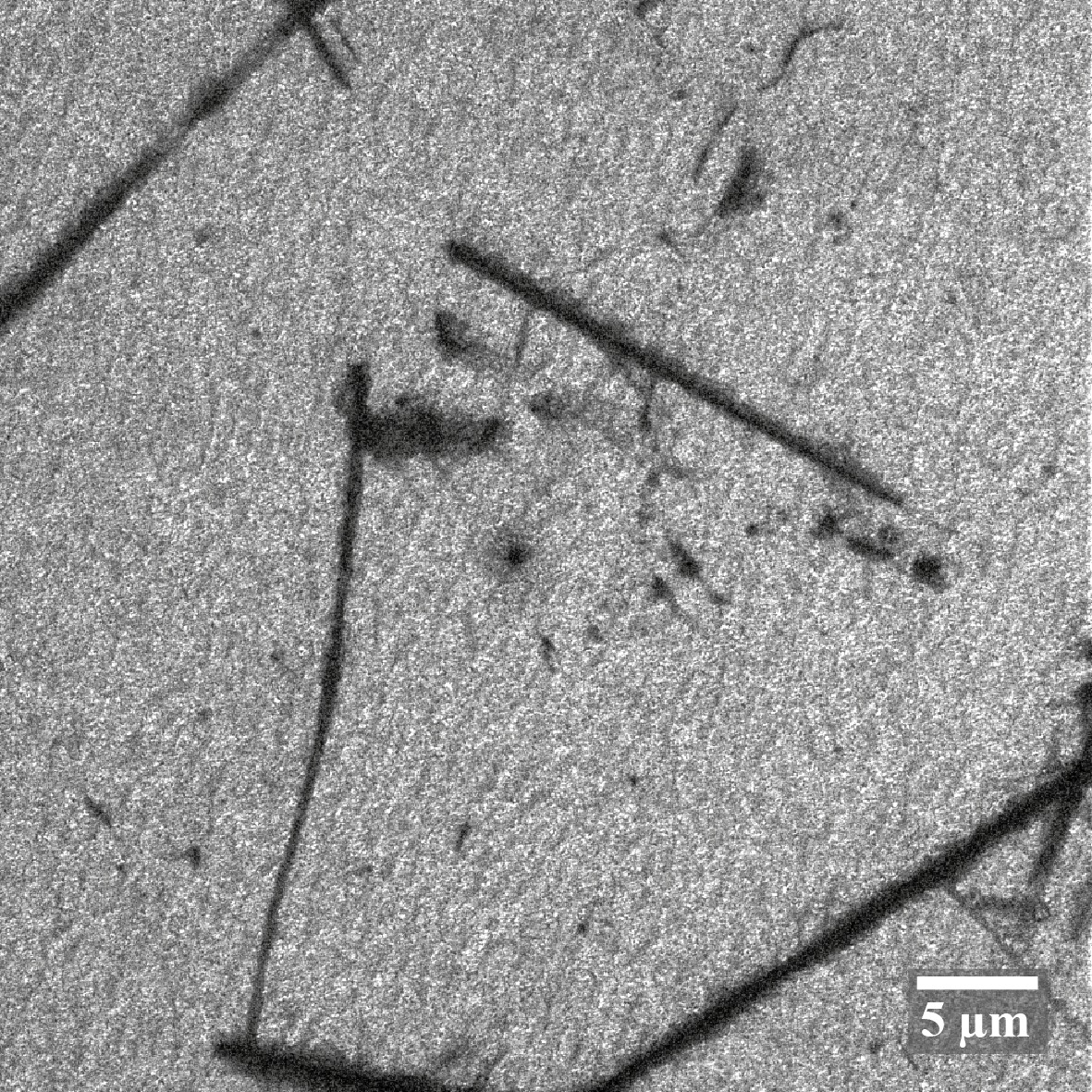}
   \caption{(IL image of a GaN surface. He$^{+}$ beam
      energy is 35\,keV. FOV: 45\,\textmu{m}}
   \label{fig:Fig_GaN}
\end{figure}

Several direct bandgap semiconductor materials were investigated in
Ref.~\lowcite{IL_Boden}, but no IL was detected. However, recently we
managed to obtain clear ionoluminescence information from such a material.
Figure~\ref{fig:Fig_GaN} shows the IL image obtained from a thin GaN film on
sapphire using a fluence of only $3\times10^{12}$\,cm$^{-2}$.
The dark lines in the IL image (Fig.~\ref{fig:Fig_GaN}) are dislocations
that are initially present in the film. These act as centers for
non--radiative recombination\cite{GaN_disl} and appear dark. The black dot
with the grey halo in the center of the image is the result of the long term
irradiation of a single pixel prior to recording the image. Due to the high
dose applied there all IL is quenched rapidly. This is the result of the
creation of various types of defects that provide non--radiative paths for
the electron de--excitation.\cite{GaN_def}

In contrast to bulk semiconductors, the authors of Ref.~\lowcite{IL_Boden} could
record IL images of agglomerates of semiconductor quantum dots. It is
suggested, that crystal defects in QDs are generated at a lower rate due to
the small size of the particles ($\approx$5\,nm). Hard nuclear collisions occur
deeper in the bulk, while in the first tens of nanometers electronic
stopping is predominant. The observed emission corresponds to the expected
band--gap transition. Nevertheless, the signal is quenched with increasing
ion fluence. 
While QD aggregates were detected with relative ease, attempts to record
IL images with higher magnification showing single QDs were not successful. 

\begin{figure}[tbp]
   \includegraphics[width=\linewidth]{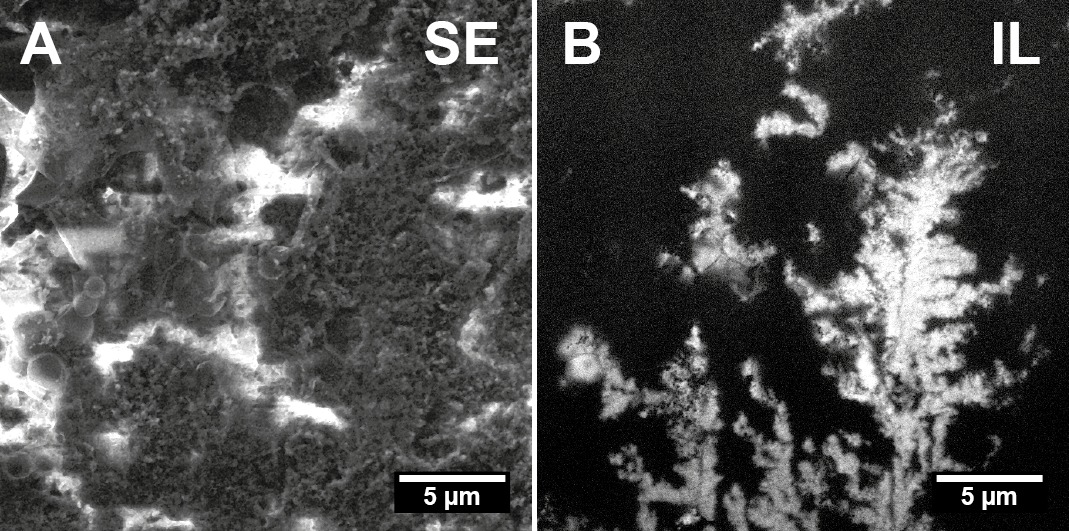}
   \caption{Simultaneously recorded SE and IL images of NaCl residues from
      water solution on a metal. IL images reveal a distribution of NaCl.
      He$^{+}$ beam energy is 35\,keV. FOV: 25\,\textmu{m}}
   \label{fig:Fig_NaCl}
\end{figure}

Alkali halides are known to exhibit intrinsic IL as a result of defect
production due to the ion beam radiation.\cite{IL_NaCl,IL_Baz,IL_NaCl_Ag}
Figure\,\ref{fig:Fig_NaCl} demonstrates
simultaneously acquired images of residues from NaCl solution on a
metal substrate. The bright areas in the SE image
(Fig.~\ref{fig:Fig_NaCl}(a)) are the aluminum substrate. The darker
areas that exhibit a rough appearance are NaCl and residue from the
drop--cast solution used. The IL image (Fig.~\ref{fig:Fig_NaCl}(b)) clearly
shows the presence of NaCl as bright ramified structures.

Ogawa et al.\cite{IL_Ogawa} have reported IL measurements on SiO$_{2}$ using
HIM. Interestingly they observe an increase of the IL intensity with
increasing fluence. The authors identify several differences between CL and
IL spectra from the same sample, but the nature of the IL signal remains
unclear.

The use of IL for analysis of biological samples is possible, but not very often
used. The advantage of IL imaging over conventional fluorescence imaging is
an absence of diffraction limits that restrict the spatial resolution. In
that respect, application of IL in HIM for bio--imaging looks promising. Few
groups have used proton beams with energies in the MeV range to investigate
biological samples such as cells\cite{Bio_Watt} and skin
tissue.\cite{Bio_Pallon} In another study several promising dyes have been
identified and can now be used to stain cells specifically for IL
investigations.\cite{Bio_Rossi} However, only limited work is done using IL
for imaging of biological samples. Using a classical ion source, a resolution
of 150\,nm has been demonstrated for ionoluminescence images in STIM
mode.\cite{Bio_Nor}

\subsection{Imaging of thin and soft layers}

\begin{figure}[tbp]
   \centering
   \includegraphics[width=\linewidth]{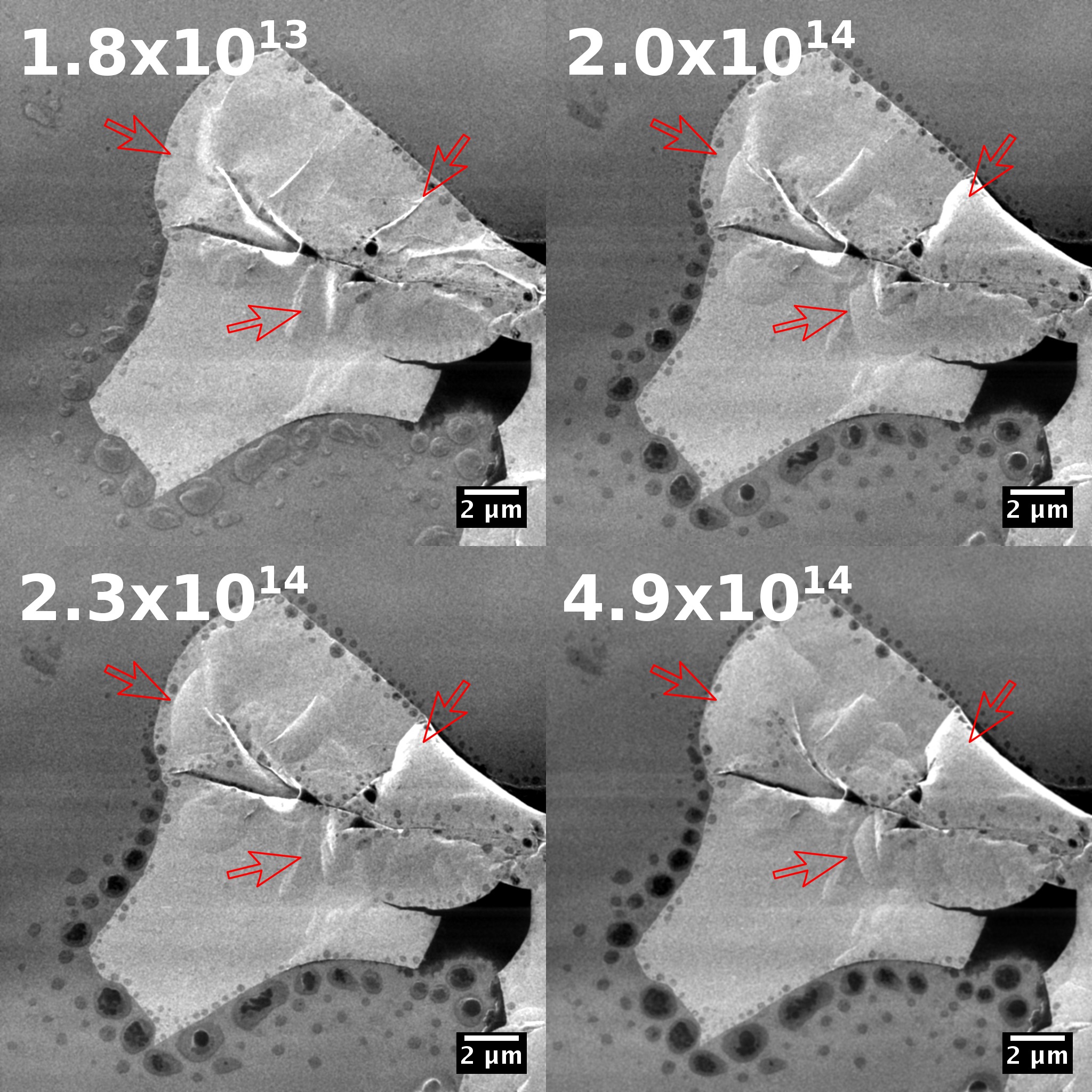}
   \caption{(Color online) Graphene flake on SiO$_2$. Using a dose of
      1.8$\times10^{13}$\,cm$^{-2}$ per image the formation of a He bubble
      can be observed. As can be seen for the bubble marked in the lower
      part of the images, holes and edges allow the He to escape and the
      bubble collapses. At a fluence of 2.0$\times10^{14}$\,cm$^{-2}$ the
      bubble is filled and it collapses two frames later after increasing
      the dose to 2.3$\times10^{14}$\,cm$^{-2}$. With further increase of
      the dose the cycle starts again. FOV is 20\,\textmu{}m. Ion doses in
      cm$^{-2}$ are indicated in the figures.}
   \label{fig:graphene-bubble}
\end{figure}

Graphene, the prototypical thin layer that currently receives a lot of
attention, has successfully been imaged by several
groups.\cite{Lemme2009,PICKARD2010,Fox2013,Bell2009,Nakaharai2013} This was
done for both suspended and free standing graphene. In
Fig.~\ref{fig:graphene-bubble} a SiO$_2$ supported graphene flake has been
imaged using mild imaging conditions.%
The flakes can be imaged easily with substantial contrast, despite their
ultra thin nature. The sequence of images presented in
Fig.~\ref{fig:graphene-bubble} shows the formation and collapse of helium
filled bubbles. Graphene has been shown to be impermeable for many gases
including helium,\cite{Bunch2008} hence the backscattered helium remains
trapped below the graphene sheet. These bubbles fill with He until either
the rim of the flake or a hole in the layer is reached. This will allow the
He to escape and the bubble then collapses. This process can be observed for
the central bubble between doses of 2.0$\times10^{14}$\,cm$^{-2}$ (initial
filling) and 2.3$\times10^{14}$\,cm$^{-2}$ (collapse and start of
refilling). The filling of the bubbles and subsequent deformation of the
graphene sheet results in the stretching of wrinkles and folds that are
present in the sheet. Assuming that the bubble has the shape of a spherical
cap with a height of 100\,nm and a diameter of 4\,\textmu{}m we can
calculate the pressure under the graphene sheet. Using the ideal gas law and
a backscatter yield of 0.006, as obtained from SRIM\cite{Ziegler2008}, we arrive at a pressure
on the order of 10\,mbar. This value agrees reasonably well with the
observed pressure range for gas bubbles under graphene on SiO$_2$ obtained
elsewhere.\cite{Bunch2008}

First principles calculations indicate that it should be possible to reveal
the graphene lattice using HIM due to minute changes in the electron
emission across the lattice.\cite{Zhang2012} However, the realization of
such an image might be difficult due to the destructive nature of the helium
beam. It is not clear if sufficient signal can be collected before the
defects will alter the local atomic configuration.\cite{Fox2013} Defects
created in graphene due to the ion--graphene interaction have been
investigated theoretically.\cite{Lehtinen2011,Ahlgren2012,Lehtinen2010}
Recent experimental investigations indicate that already at very low doses
severe damage is done to the graphene sheet and many defects are
created.\cite{Fox2013} Safe imaging doses for suspended graphene seem to be
as low as 10$^{13}$ to 10$^{14}$\,cm$^{-2}$.\cite{Fox2013} For supported
graphene, as shown above, the critical dose is certainly lower due to the
more destructive nature of recoils from the support material.

An exceptional imaging result has been achieved for carbon nano membranes
(CNM).\cite{Turchanin2012} Such free standing membranes, that are formed by cross
linking self--assembled monolayers and then removing the substrate, are
difficult to image using conventional SEM. However, in HIM these
CNMs are not only visible, but morphological details such as folds and
nanometer sized holes created by slow highly charged ions, are also
visible.\cite{Ritter2013}

Given the nature of ultra thin films and the sampling volume of the
available secondary imaging particles in HIM, the SEs are the obvious choice
for imaging such samples. An example of the successful application of HIM
to soft layers is the visualization of the phase separation in a mixed
poly(3-hexylthiophene)/[6,6]-phenyl-C$_{61}$-butric acid methyl ester
(P3HT/PCBM).\cite{Pearson2011} Such blends are typically used in organic
photovoltaic solar cells and represent an important materials class.
Although both molecules are essentially carbon, the different number of
$\pi$ and $\sigma$ bonds in the two polymers results in minute changes of
the SE yield.\cite{Kishimoto1990} An other example is the clear
identification of self--assembled monolayers (SAM) deposited onto
SiO$_2$.\cite{Hlawacek2012,George2012} Here, differences in work function
allow the identification of SAM molecules and give rise to the observed
contrast.

A delicate method of visualizing non--continuous ultra thin layers and their
different properties utilizes BSHe. For a substrate oriented in a channeling
condition, a very low BSHe signal is recorded due to the extended range of
He along the low index channeling direction.\cite{Veligura2012} However, at
places where a thin surface layer with different atomic positions is
present, scattering will occur and an increased BSHe signal can be observed.
Although the relative mass of the elements in the adlayers and the
underlying bulk are important, this effect works for any combination of bulk
and adlayer elements.\cite{Mocking2012} In particular it works for light
elements on a heavy substrate.\cite{Hlawacek2012,George2012}

\subsection{Voltage contrast}

Similar to SEM, local electromagnetic fields in the sample will influence the
yield, angular distribution and energy of the generated SE. This can be
utilized to image dopant distribution\cite{Jepson2009}
and electronic potential distribution. A similar application utilizes static
capacitive\cite{Ura2000} contrast to reveal conducting features buried below an
insulating cover.\cite{Scipioni2009a}

\section{Materials modification}

The availability of different gases for GFIS---such as
Neon\cite{Rahman2012,Livengood2011}---and the combination with a classic
liquid metal Gallium Focused Ion Beam (FIB)\cite{Notte2012,Wu2013} makes the
technique interesting for various types of materials modifications. The fine
beam produced by the GFIS has the potential to engineer structures with a
length scale that is well below what is currently possible with LMIS based
FIB techniques. Contrary to e--beam writing, which can achieve similar
critical lengths, the removal of material is also
possible.

\subsection{Resist patterning}

\GH{Writing structures into resists
is usually done using gas injection systems (GIS) in FIB or, if higher
resolutions are needed, by using e--beam lithography.} In particular
the latter suffers from the so--called proximity effect.\cite{Chang1975}
The deposition parameters necessary to achieve a constant feature size will
depend on the distance to the next feature in the proximity of the beam.
This effect is a consequence of the relatively large lateral range of the
electrons in the resist material. 
Structures produced by low mass ion beams are less sensitive to this effect.\cite{Itakura1985} The
near complete absence of this effect is related to the fact that in the
surface near region there is practically no scattering when using ion beams
(see also Fig.~\ref{fig:charged-part-beams}).
Furthermore, very little backscattering
occurs in the usually relatively light resist materials. In addition 
the very low energy SE
(<10\,eV)\cite{Petrov2011,Ramachandra2009,Ohya2009,Petrov2010} in HIM are
localized close to the beam path, whereas the SEs in e--beam lithography
have a higher energy (>100\,eV) and therefore travel relatively long
distances in the resist material.\cite{Seah1979} As a consequence resist
modification by ions happens only in a very small volume along the beam
path. An extensive overview comparing electrons and ions for their useful
application in lithography is given by Melngailis.\cite{Melngailis1993}

With the availability of He GFIS new standards have been set in terms of
achievable line
width.\cite{Winston2009,Maas2010,VanderDrift2011,Alkemade2012} This has at
least partially been made possible by the above discussed, near complete,
absence of the proximity effect when using a highly focused He beam.
\GH{However, to date resists initially developed for e--beam lithography such as
hydrogen sylsesquioxane (HSQ) and polymethyl methacrylate resist (PMMA) are used. New resist materials optimized for the large amount
of energy deposited by the ion beam could potentially lead to even better
results.}

\subsection{\GH{Beam induced deposition}}

\begin{figure}[tbp]
   \centering
   \includegraphics[width=\linewidth]{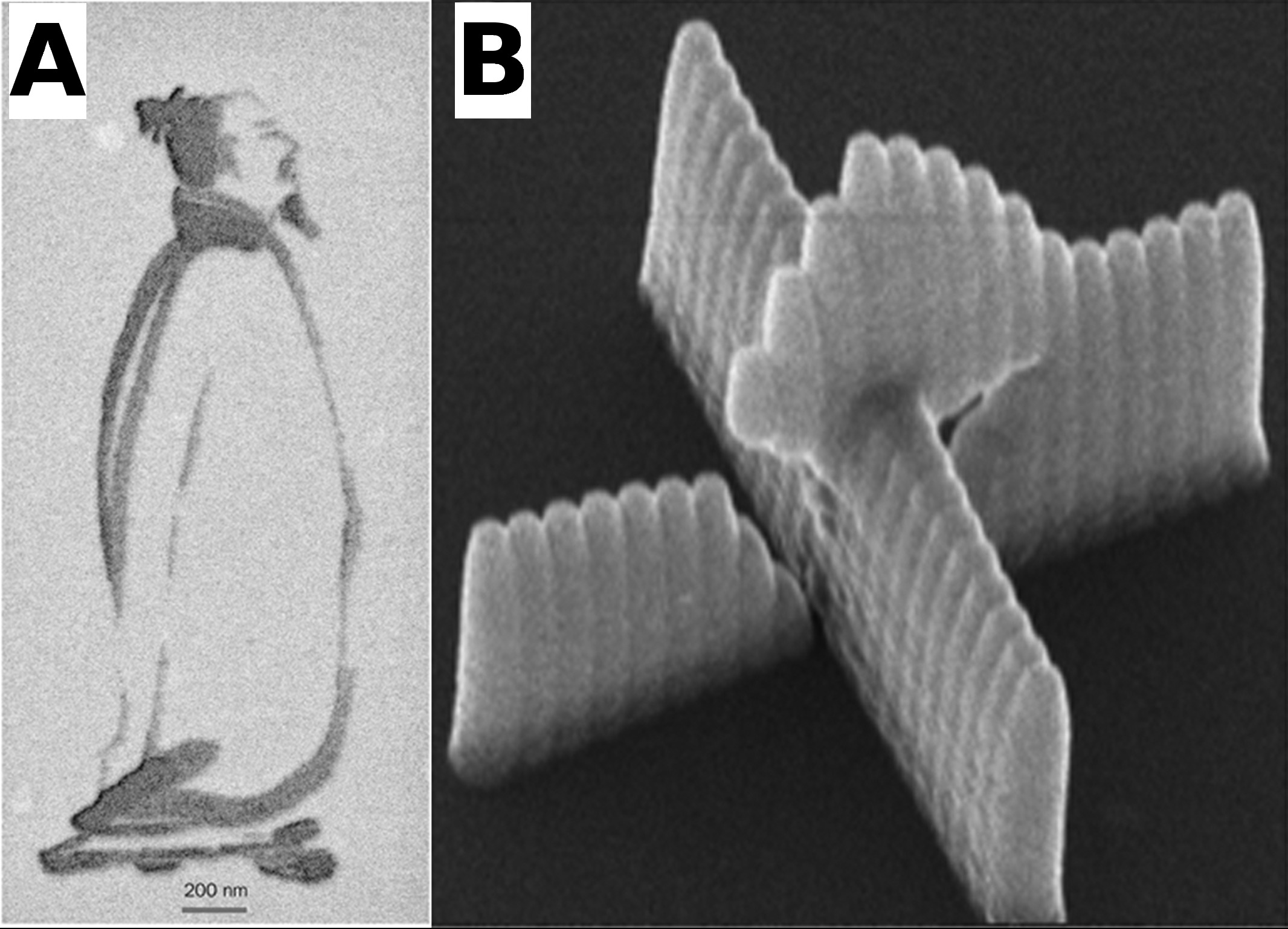}
   \caption{Examples of structures created using He beam induced metal
   deposition. (A) 100.000$\times$ demagnified copy of a 13th century
   painting of the Chinese poet Li Bai. (B) Stacked cross formed by
   exploiting shadowing of the precursor gas (FOV: 1\textmu{}m).
   \citetext{Adapted with permission from Scanning 34, 90--100
   (\citeyear{Alkemade2012}). Copyright 2012 Wiley Periodicals, Inc.}}
   \label{fig:metal-deposition}
\end{figure}

Several precursor materials have been explored including MeCpPtMe$_3$, TEOS,
and
XeF$_2$.\cite{Chen2010a,Maas2010,Scipioni2011,Alkemade2012b,Sanford2009a}
Recently, also a tungsten based metal precursors has successfully been
used.\cite{Kohama2013} With this extremely flexible method a wide range of
features can be directly sculpted including 3D
structures.
Figure~\ref{fig:metal-deposition} shows examples of structures created using
Ion Beam Induced Deposition (IBID) in the group of P. Alkemade.\cite{Alkemade2012}
However, a detailed understanding of the deposition process is needed to
unleash the full potential of the method for resist patterning. A
model has been proposed that identifies primary ions and secondary electrons
as being responsible for the vertical growth of the structures. Scattered
ions and recoil atoms on the other hand are responsible for the lateral
growth of the structures.\cite{Alkemade2010} Simulation software
(\textit{EnvisION}) based on this model is available and can be used to
predict the metal deposition process in HIM.\cite{Chen2010a} The program is
based on available concepts and code of an electron--beam--induced
simulation\cite{Smith2008} and the \textit{IONiSE}\cite{Ramachandra2009} Monte
Carlo program. Experimental confirmation of results showed the applicability
of the simulation method.\cite{Chen2010a}

\subsection{Direct write lithography}

Different from what has been discussed in the previous section, here we will discuss
structuring results that do rely on the sputtering of sample atoms.
Sputtering is a well known phenomenon that has been extensively studied.\cite{Sigmund1987,Thompson1981,Behrisch2007} However,
\GH{with only a few exceptions} these works usually
focus on ions of lower energies of only a few hundred eV to a
few keV for the relevant gases Helium and Neon. \GH{For higher energy ranges the bombardment species is either
   not a noble gas or generally a heavier element.} The
energy range of interest for HIM is not so well investigated. 
Surprisingly, many as yet unexplained phenomena, such as very high and
fluence dependent sputter yields are observed.\cite{Fowley2013} Due to the
sub nanometer size and the collimated nature of the HIM ion beam, it is
possible to mill features with very small critical dimensions not possible
with current FIB technology. We start by discussing results obtained
on the most simple milled structure---a pore. Such pores are successfully
used for biomolecule detection. The achievable diameter can be as small as
4\,nm, which is 60\% smaller compared to other one step milling
techniques.\cite{Yang2011} Most importantly the quality of these pores for
biomolecule detection is comparable to state--of--the--art pores fabricated
using TEM\cite{Storm2005} or FIB.\cite{Li2001} However, these techniques
create smallest pores by closing existing bigger pores in a controlled way.
\GH{The method relies on the fluidization of the material under 
irradiation and subsequent deformation which is driven by the surface
tension and diffusion of ad--atoms created by an ion beam.\cite{Storm2005,Li2001}
Although it has not been demonstrated to date, this should also be possible by 
careful implantation of He into the area surrounding the pore. In addition to
the effects observed elsewhere,\cite{Storm2005,Li2001} an additional volume increase and subsequent reduction of
pore size can occur due to the formation of nano-sized He
bubbles.\cite{Veligura2013a,Fox2012}} Substantial
redistribution of unsputtered material has already been observed when
milling bigger pores.\cite{Marshall2012} As a result, comparable diameters
should be achievable.

\begin{figure}[tbp]
   \centering
   \includegraphics[width=\linewidth]{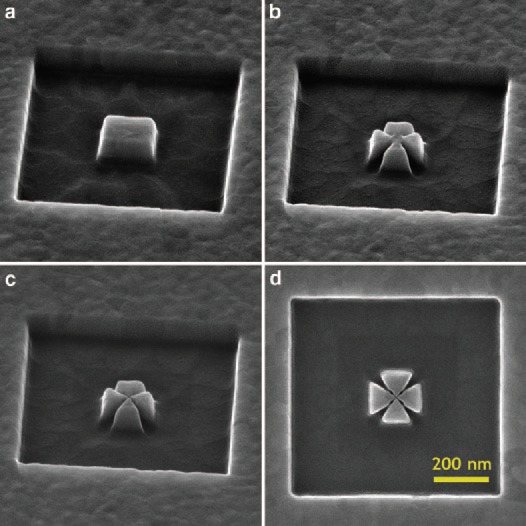}
   \caption{(Color online) Milling a gold on glass substrate over multiple length scales. (a) The
      initial structure is milled using a Gallium beam. (b) Intermediate
      features are milled using a GFIS Neon beam. (c) Final milling of details is
      performed with a He ion beam. (d) Achievable feature sizes are as small as
      13\,nm. \citetext{Reproduced with permission from Micros. Today 20,
      16--22 (\citeyear{Notte2012}). Copyright 2012 Microscopy Society of
   America.}}
   \label{fig:triple-milling}
\end{figure}

Recently a GFIS using Neon\cite{Tan2010} became available. Its combination
with a classical liquid metal Gallium FIB column allows for efficient
milling of structures at several length scales.\cite{Notte2012}
Figure~\ref{fig:triple-milling} shows a structure milled into gold using all
three beams. While initial cut outs are made using the LMIS gallium source
(Fig.~\ref{fig:triple-milling}(a)) finer details are cut using the GFIS and
Neon (Fig.~\ref{fig:triple-milling}(b)). Smallest details with a length
scale of the order of 10\,nm are patterned using helium and the GFIS
(Fig.~\ref{fig:triple-milling}(c)). The same gas source combination is also
used for imaging of the structures.

Another exciting application of helium ion beams for direct write
lithography is the successful preparation of devices based on
graphene.\cite{Boden2011} The possibility to precisely cut graphene
with nanometer precision has been demonstrated earlier by Lemme et
al.\cite{Lemme2009} Unfortunately, the unique properties of graphene are
very sensitive to damage introduced by the impinging
ions.\cite{Lucchese2010} Consequently, only very small fluences of the order
of $10^{13}$\,cm$^{-2}$ can be used to image graphene without damaging it to
the extent that the unique properties are lost.\cite{Fox2013}
Unfortunately, this excludes high resolution imaging of supported as well as
free standing graphene.

\subsection{Helium implantation and defect engineering}

Helium implantation is unavoidable in HIM. It is therefore important to
understand the associated defect structures, their evolution and response
to external influences. It has to be made clear that the phenomena discussed
in this section appear at doses higher than those normally used in HIM
imaging. However, special applications (BSHe or IL imaging) may require
relatively high doses. Obviously, He implantation plays an important role
when the HIM is used for materials modification, as discussed in the previous
section. \GH{A dramatic example is shown in Ref.~\lowcite{Wang2012a} in the
context of the fabrication of domain wall magnetoresistance devices.
Although successful fabrication of domain wall magnetoresistance devices was
demonstrated at lower doses, at higher doses the implanted He leads to a
swelling and surface deformation that in turn results in the destruction of the
patterned devices.}

\begin{figure}[tbp] 
   \centering 
   \includegraphics[width=\linewidth]{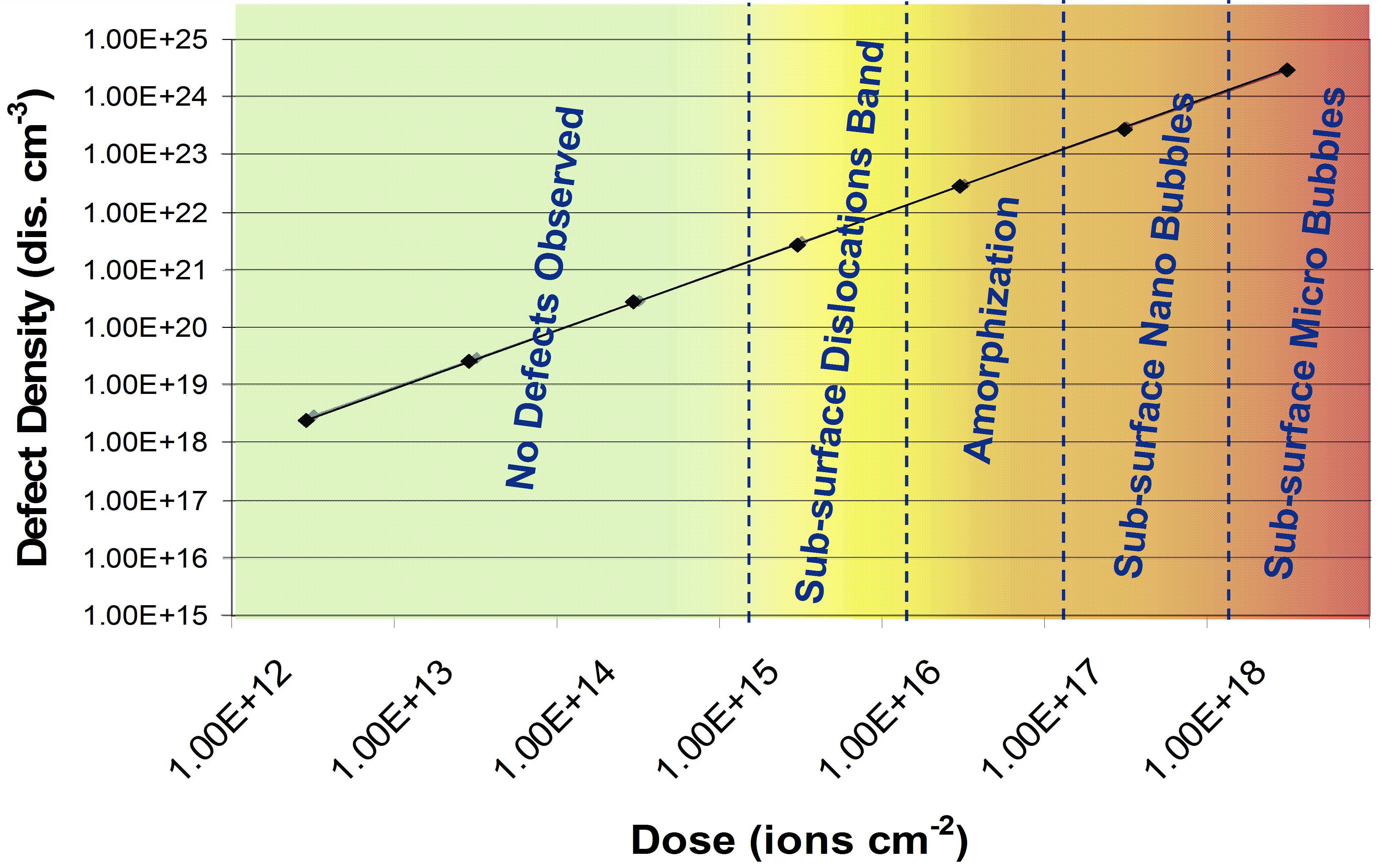}
   \caption{(Color online) Defect density versus He ion dose for silicon and copper. The
   regimes in which specific damage mechanisms dominate are marked.
   \citetext{Reproduced with permission from J. Vac. Sci. Technol. B
27, 3244 (\citeyear{Livengood2009}). Copyright 2009 American Vacuum
Society.}}
   \label{fig:damage-diagram} 
\end{figure}

Initial studies that allow the comparison of experimental damage volumes
with SRIM were made by Livengood et al.\cite{Livengood2009} One of the
results in this publication is summarized in an overview over the different
damage regimes that are to be expected for a wide range of He doses (see
Fig.~\ref{fig:damage-diagram}).
Interestingly, this diagram is valid for both semiconductor materials
(silicon) and metals (copper). The validity of these damage regimes has been
demonstrated by several groups. 
A nice visualization of the bubbles that form at depths comparable to the
range of the ions in the material is shown by Ref.~\lowcite{Fox2012}.
The pressure inside these bubbles and blisters is high. 
For the rather soft and ductile metal gold,
pressures of up to several GPa can be reached in the initial He
nano--bubbles.\cite{Veligura2013a} These bubbles form at open volume defects
that are either present in
the material or are created by the beam.\cite{Laakmann1987,Rajainmaki1988} 
The growth of these initial, nano--sized, He bubbles is independent of the
primary energy and leads to observable materials modification at fluences in
the low 10$^{17}$\,cm$^{-2}$ range.\cite{Livengood2009,Veligura2013a}
However, with increasing He dose these bubbles start to coalesce, provided
they are formed deep enough in the sample to not immediately reach the
sample surface. What follows is a rapid expansion of a blister on the
surface. The pressure inside such a blister of a few hundred
MPa\cite{Veligura2013a,Wang2012a} is substantially lower than in the initial
nano--sized bubbles. The effective formation and rapid growth of He induced
blisters is at least partly due to the fact that highly compressed He gas is
an ideal stopping medium for He. This results in an amplification of the
damage in the vicinity of the bubble and the observed accelerated growth.

\section{Summary}

In the past half decade Helium Ion Microscopy has proven to be an interesting
alternative to its direct and well established competitors: scanning electron
microscopy and gallium focused ion beam. However, the authors strongly
believe, and hope to have demonstrated above, that HIM is more than a
replacement for SEM and FIB. It excels over SEM and FIB in particular for
\begin{itemize}
   \setlength{\itemsep}{1pt}
   \setlength{\parskip}{0pt}
   \setlength{\parsep}{0pt}
   \item high resolution imaging of uncoated biological samples
   \item imaging of insulating samples
   \item a high surface sensitivity
   \item imaging of ultra thin layers
   \item materials modification with unprecedented resolution
   \item direct write lithography
   \item resist patterning
\end{itemize}

At the same time new applications and techniques are constantly being
developed and refined. In particular, analytical additions are needed and
are currently being developed. While ionoluminescence has its special
applications, additions such as secondary ion mass
spectroscopy\cite{Wirtz2012} still need to prove their applicability in real
materials science problems. However, although spectroscopy is currently
still in its infancy, the extension of the technique to other gases such as
neon, and potentially even heavier ones, guarantees an exciting future of
the technique in particular for nano--fabrication applications.

\begin{acknowledgments}
   This research is supported by the Dutch Technology Foundation STW, which is
   the applied science division of NWO, and the Technology Programme of the
   Ministry of Economic Affairs.
\end{acknowledgments}


%

\end{document}